\documentclass[preprint,superscriptaddress,amsmath,amssymb,aps]{revtex4-2}

\usepackage{graphicx}
\usepackage{dcolumn}
\usepackage{bm}

\usepackage{mathrsfs}
\usepackage[mathcal]{eucal}
\usepackage{mathtools}

\begin{document}


\title{Stability of bond clusters with a characteristic length scale for load distribution}


\author{Yannick L{\"u}demann}
\affiliation{University of G\"ottingen, Institute for the Dynamics of Complex Systems,  Friedrich-Hund-Platz 1, 37077 G\"ottingen, Germany}

\author{Stefan Klumpp}
\affiliation{University of G\"ottingen, Institute for the Dynamics of Complex Systems,  Friedrich-Hund-Platz 1, 37077 G\"ottingen, Germany}

\author{Komal Bhattacharyya}
\thanks{Correspondence to stefan.klumpp@phys.uni-goettingen.de (SK) or komal.bhattacharyya@uni-goettingen.de (KB).}
\affiliation{University of G\"ottingen, Institute for the Dynamics of Complex Systems,  Friedrich-Hund-Platz 1, 37077 G\"ottingen, Germany}


\begin{abstract}
In biological materials, strong binding despite an applied load force is often based on clusters of dynamic bonds that share the load. Different macroscopic behaviors have been described depending on whether the load is shared locally or globally in the force-depended unbinding rate. Here we introduce and study a model in which the load is distributed over a characteristic length scale, introduced by an exponential decay. The model contains the local and global scenario as limiting cases and smoothly interpolates between them. We derive approximations in which some analytical results can be obtained. In particular, we derive rupture conditions and validate these with stochastic simulations. The model shows two main pathways for failure of the bond cluster, due to rupture of all bonds above a critical force and due to the formation of a critical crack, a large gap between closed bonds that spreads in both directions.
\end{abstract}
\maketitle


\newpage


\section{Introduction}
Biological systems respond rapidly to mechanical stimuli due to molecular bonds that are both dynamic and highly sensitive to applied forces. Specific interactions between cell-surface receptors and complementary ligands on neighboring cells or the extracellular matrix underlie the mechanosensitivity of cell–cell and cell–matrix junctions \cite{gao_probing_2011}. These bonds can stochastically bind and unbind under mechanical load \cite{evans1997dynamic,bell1978models}.
To understand how collections of such bonds respond to force, minimal one-dimensional (1D) models have been widely employed. These models capture essential features of cell–matrix adhesions, and have been used to explore the stability of bond clusters through both analytical theory and stochastic simulations \cite{erdmann2004adhesion, erdmann2004stability,seifert_dynamic_2002,seifert_rupture_2000}.
 
Beyond adhesion complexes, transient binding also governs the behavior of cytoskeletal networks \cite{pegoraro2017mechanical,blanchoin2014actin}. Crosslinking proteins such as $\alpha$-actinin, filamin, and fascin dynamically bind semi-flexible actin filaments \cite{blanchoin2014actin}. Arrays of closely spaced, reversible crosslinkers are thought to underlie the power-law frequency dependence of the storage and loss moduli observed in reconstituted cytoskeletal networks \cite{broedersz_cross-link-governed_2010, muller_rheology_2014}. Likewise, the transport of vesicular cargo along cytoskeletal filaments depends on the dynamic attachment of the cargo to the filament via molecular motors \cite{klumpp2005cooperative,mclaughlin2016collective}.
  
 In such models—whether modeling adhesion or cytoskeleton—the rupture time (i.e., the time until all bonds unbind) serves as a key mechanical observable. Theoretical studies have identified distinct scaling regimes for the rupture time as a function of system size, rebinding rate, and applied force \cite{seifert_dynamic_2002}. 
Stability is enhanced by rebinding of open bonds
\cite{erdmann2004stability,erdmann2004adhesion}, but reduced when crosslinkers are mobile and can rebind at different locations \cite{mulla2018crosslinker}. 
Bistability has also been observed in models with position-dependent binding rates \cite{erdmann_bistability_2006}. 

When an external force — either constant \cite{erdmann2004stability} or increasing over time \cite{evans1997dynamic,erdmann2004adhesion} — is applied, it is typically assumed to be equally distributed among all closed bonds. 
Due to load sharing among bonds, it is the collective behavior of clusters rather than individual bonds that governs rupture dynamics \cite{erdmann2004adhesion}. 
However, in realistic semi-flexible filament networks such as the cytoskeletal networks, force transmission is spatially inhomogeneous. How a force is distributed depends on many factors: filament bending rigidity, network connectivity, and the density of crosslinkers \cite{broedersz2011criticality, alvarado2017force}.  Indeed, the low-frequency plateau of the storage modulus of entangled \cite{morse_viscoelasticity_1998,broedersz_modeling_2014} and permanently crosslinked networks \cite{shin_relating_2004} depends on typical length scales such as the persistence length of the filaments, the mesh size, the entanglement length \cite{morse_viscoelasticity_1998} of the network, and the average spacing between crosslinkers \cite{shin_relating_2004}. All the effects of the network can be reduced into a length scale that determines the force distribution in the 1d model of bond clusters. So, it is especially interesting to study how the system behaves depending on how the forces applied to this minimal model are being distributed over the system. 
Previous work has shown that when an external force is distributed only to the nearest closed bonds instead of all closed bonds, the rupture time is reduced and rupture is initiated at a crack or a big cluster of open bonds \cite{mulla2018crack}. Likewise, unequal force sharing has been discussed extensively in the context of molecular motors, where unequal force sharing may arise from the stochastic stepping of the motors and reduces the benefits of motor cooperativity \cite{mclaughlin2016collective,berger2012distinct,kunwar2011mechanical,yadav2022sliding}. However, a general understanding of the influence of the force distribution length scale on the mechanical response of such systems is still lacking. 

 In this work, we introduce a general model of a one-dimensional ensemble of bonds with force-dependent unbinding kinetics, where an external force is distributed among the bonds in a fashion that shows an exponential with the distance from the attack point of the force. The length scale of this decay, which may correspond to any of the length scales mentioned above, provides a control parameter for the extent of coupling of the bonds via the sharing of force. The model includes a limiting cases the local \cite{bell1978models,mulla2018crack} and global \cite{mulla2018crack} models studied earlier. We further introduce an approximation based on a continuous force distribution to simplify our model further in order to obtain analytical results and theoretical insights regarding the bond dynamics. Specifically, we derive rupture conditions, i.e. conditions for critical values of the parameters that will lead to rupture. These are found to be in good agreement with stochastic simulations of our full model. We observe two rupture pathways, via a spreading of a crack and via uniform bond opening, which are dominant in the local and global limit of the model and coexist for intermediate force distribution lengths.

\section{A model for bond dynamics with nonlocal coupling}
\label{sec:model}

\begin{figure}[tb]
    \centering
    \includegraphics[width=0.9\textwidth]{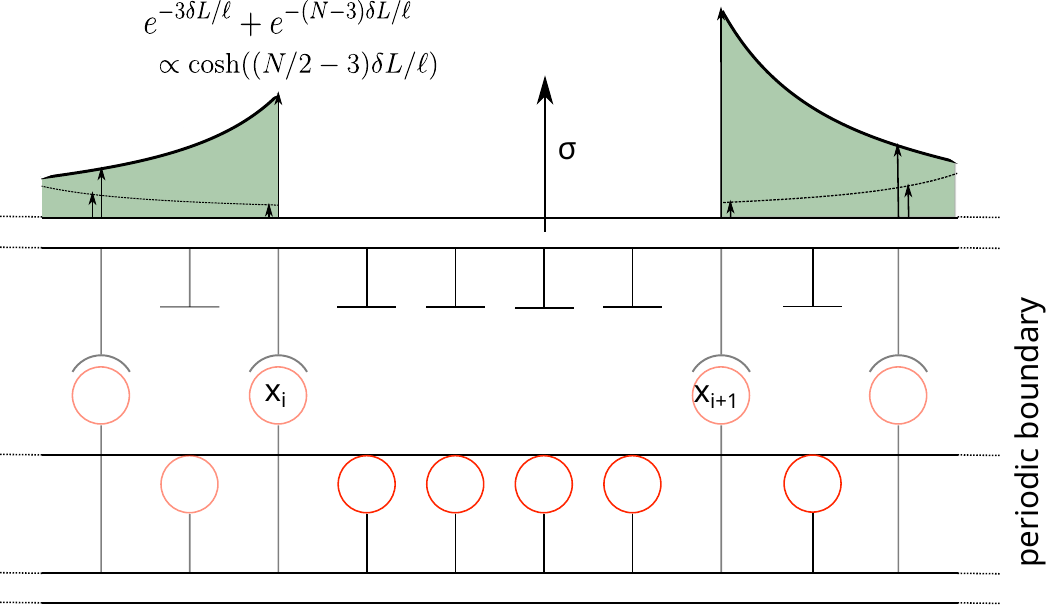}
    \caption[Model of force distribution among parallel bonds]{Model of force distribution among parallel bonds: The force $\sigma$ attacking at the third bond position in the gap is redistributed to the surrounding bonds as follows. The portion of force on each bond is proportional to an exponential function in the distance from the site under consideration, resulting in a hyperbolic cosine. This resulting force is indicated here  for the left neighbour.}\label{fig:force distribution schematic}
\end{figure}

We consider a stochastic model for $N$ bonds, equally spaced over a length $L$, which can be either open or closed, see Fig.~\ref{fig:force distribution schematic}. The distance between two bond sites is $\delta L$, so that $N \delta L = L$. We denote by $n$ the number 
of closed bonds at a particular time and label their positions $x_i$, where $i \in \{0,1,\dots,n-1\}$. 
We assume periodic boundary conditions at the two ends so that without loss of generality we may set the position of the first closed bond at $x_0=0$, which can always be achieved by relabeling the bonds. We will occasionally write $x_n=L$ in order to keep summation over all bonds simple; however, $x_n$ denotes the position of the same bond as $x_0$. The rate $k_\text{on}$ of a particular open bond closing is taken to be constant whereas the rate $k_\text{off}$ of a closed bond opening depends exponentially on the force acting on that bond. We express the total force $F$ acting on the ensemble of bonds in terms of a force density per length, $\sigma=F/L$, corresponding to an average force $\sigma\, \delta L$ per bond (open and closed). 

The defining feature distinguishing different models is how the total force is distributed among the closed bonds, i.e., how the force $f_i$ on the closed bond $i$ depends on $\sigma$. For example, the classical Bell-Evans model distributes the force equally, so the force on each bond is $f_i=F/n=\sigma L/n$ \cite{bell1978models,evans1997dynamic}. 
A local model has also been proposed, where the force on a bond $i$ depends on the distance $l_i$ between the neighboring closed bonds on both sides and $f_i=\sigma l_i/2$ \cite{mulla2018crack}. 

Here we introduce a general model that includes both these scenarios as limiting cases and that is parametrized by a decay length $\ell$ of the force distribution. 
Thus, the off-rate of bond $i$ increases exponentially as
\begin{equation}\label{eq:general off rate}
    k_\text{off}(\sigma, \ell, i) = k_\text{off,0}\exp\Big(f_i(\sigma, \ell )/f_{d}\Big),
\end{equation}
with a force $f_i$ that depends on the force density $\sigma$, the decay length $\ell$ as well as possibly the entire configuration of bonds. $f_d$ is a characteristic force scale, the detachment force. We propose the following definition for $f_i(\sigma, \ell )$: As $\sigma$ is the applied force per unit length, we assume that a force term $\sigma \delta L$ attacks at every site of a bond (open or closed) and is distributed to all closed bonds as follows. We will denote the force that reaches the closed bond at $x_i$ for $i\in\{0,\dots,n-1\}$ from site $j\in\{0,\dots,N-1\}$ by $f_{ij}$. Since the forces that stem from position $j$ have to add up to $\sigma\, \delta L$, we have,
\begin{equation}\label{eq:normalization force}
    \sum_{i=0}^{n-1}f_{ij} = \sigma\, \delta L.
\end{equation}

We assume a force distribution that decays exponentially with distance from the attack point of the force. More precisely, consider the external force $\sigma\, \delta L$ that is applied at bond site $j$. If $d_{ij}$ is the distance between the closed bond $i$ at position $x_i$ and the site $j$ where the force attacks, the portion of the force that is carried by the closed bond at $x_i$ should be proportional to $\exp\big(-d_{ij}/\ell \big)$ (or $\exp\big(-(L-d_{ij})/\ell \big)$, respectively, due to the periodic boundary condition). Adding up all these contributions while also keeping the periodic boundary conditions in mind gives
\begin{equation*}
\begin{aligned}
    \sum_{k=0}^\infty \exp &\bigg(-(d_{ij}+k\cdot L)/\ell\bigg) + \exp\bigg(-(L-d_{ij}+k\cdot L)/\ell\bigg)\\
    =&\bigg(\sum_{k=0}^\infty \exp\big(-L/\ell\big)^k\bigg)\cdot\Big(\exp(-d_{ij}/\ell) +\exp(-(L-d_{ij})/\ell)\Big)\\
    =& \frac{2\exp(-L/2\ell)}{1-\exp(-L/\ell)}\cdot\cosh\Bigg(\Big(\frac{L}{2}-d_{ij}\Big)/\ell\Bigg).
\end{aligned}
\end{equation*}
The prefactor is independent of $i$ and $j$ and can be absorbed into the normalization constant $C_j$, so that indeed 
\begin{equation}\label{eq:full_Model_general}
    f_{ij} = C_j\cdot\cosh\left(\left(\frac{L}{2}-d_{ij}\right)/\ell\right)
\end{equation}
with the normalization constant 
\begin{equation}
    C_j = \frac{\sigma\ \delta L}{\sum_{q=0}^{n-1}\cosh\Big((\frac{L}{2}-d_{qj})/\ell\Big)}
\end{equation}
to satisfy Eq.~\eqref{eq:normalization force}.
In Fig.~\ref{fig:force distribution schematic}, we present a visualisation of how the force on each bond is obtained in this model as the sum of combinations of two exponential terms. The force term in Eq.~\eqref{eq:general off rate} is then computed as 
\begin{equation}\label{eq:general force}
    f_i = \sum_{j=0}^{N-1}f_{ij}= \sum_{j=0}^{N-1}C_j\cdot\cosh\left(\left(\frac{L}{2}-d_{ij}\right)/\ell\right).
\end{equation}

In the limiting cases where the decay length of the force is either very large or very small, $\ell\rightarrow \infty$ and $\ell \rightarrow 0$, our model reduces to the global and local model of ref. \cite{mulla2018crack}, respectively: For $\ell\rightarrow \infty$, we obtain
$\cosh\Big((\frac{L}{2}-d_{ij})/\ell\Big)\rightarrow 1$ for all $i$ and $j$ and the normalization condition \eqref{eq:normalization force} gives $C_j = \sigma \delta L/n$ for all $j$. As a consequence, 
\begin{equation}\label{eq:glocal model forces}
    f_i=
    \sum_{j=0}^{N-1} \frac{\sigma\ \delta L}{n}\cdot1 = \frac{\sigma L}{n} ,
\end{equation}
as in the global model. 

For $\ell\rightarrow 0$, we can make the following approximation:
\begin{equation}\label{eq:cosh approximation}
\begin{aligned}
    \cosh\Big((\frac{L}{2}-d_{ij})/\ell\Big) &= \frac{1}{2}\bigg(\exp\Big(\Big(\frac{L}{2}-d_{ij}\Big)/\ell\Big)+\exp\Big(-\Big(\frac{L}{2}-d_{ij}\Big)/\ell\Big)\bigg)\nonumber\\
    &\approx \frac{1}{2}\exp\Big(\Big|\frac{L}{2}-d_{ij}\Big|/\ell\Big).
\end{aligned}
\end{equation} 
For given $j$, the biggest portion of force applied at that site is diverted to those bonds $i$ that maximize the argument $|L/2-d_{ij}|/\ell$ of the exponential, which are the ones that are closest to the site $j$. In fact, since for small $\ell$ the exponential function in Eq.~\eqref{eq:cosh approximation} is very steep, either the closest bond carries the entire force $\sigma \delta L$ or the nearest neighbours to either side, if they are at the same distance, carry a force $\sigma \delta L/2$ each. In particular, the force that is applied to all the sites of the gap between two neighbouring closed bonds is evenly split between both as half of the open bonds in-between are closer to each closed bond respectively and in case of an odd number, the force on the middle bond is also shared evenly. This leads to the expression for local load sharing,
\begin{equation}
    f_i  = \frac{\sigma l_i}{2},
\end{equation}
with the gap size $l_i$ between the nearest-neighbours of the $i$th bond, that is $l_i = x_{i+1}-x_{i-1}$ for $i\in\{1,\dots,n-1\}$ as proposed in ref.~\cite{mulla2018crack}.

\section{Approximation considering a continuous distribution of force}
\label{sec:continuum_approx}
The generalized model introduced in the previous section can describe realistic distributions of the external forces among the closed bonds more accurately than the local or global model, however the fact that every closed bond is directly influenced by every other one also makes it computationally more involved. Accordingly, we approximate Eq.~\eqref{eq:general force} with a simplified form, essential for interpreting the model. As a first step, we will approximate the summation over all bonds by an integration, i.e., instead of adding the forces applied to the discrete sites of the bonds, we integrate the force density over the length of the system.  As we will see, this reduces the sum over all bonds to a sum over all closed bonds.

\begin{figure}[tb]
    \centering
    \includegraphics[width=0.8\textwidth]{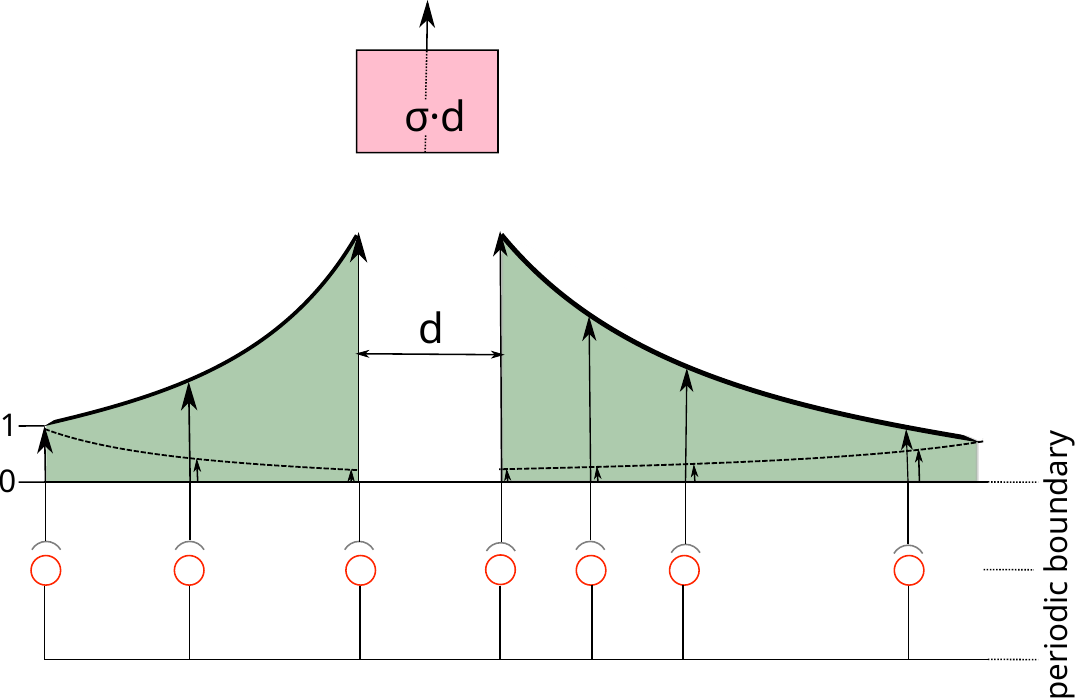}
    \caption[Illustration of the approximated force distribution with continuously attacking external force]{Illustration of the approximated force distribution: The force $\sigma\cdot d$ that attacks at a gap of length $d$ is split up equally between right and left. The sums of exponentials in Eq.~\eqref{eq:force_approximation} describe an exponential decay of the force with the distance to the gap. The fractions correspond to the portion carried by the bond at $0$ since its exponential is simply $1$. The force $f_0$ is attained by summation over all gaps.}
    \label{fig:continuous force approximation}
\end{figure}

$\sigma$ is the force density that acts continuously on the length of the system. This means that we can consider a force that acts on a closed bond at $x_i$, $i=0,1\dots, n-1$ not only from the discrete site $j \in \{0,\dots ,N-1\}$, but from any position $y$ in $[0,L]$, 
\begin{equation}\label{eq:force_from_to}
    f_{x_i,y} = \sigma\cdot c_y\cdot\cosh\left(\frac{L/2-|x_i-y|}{\ell}\right) 
\end{equation}
Here $c_y$ is the normalization constant so that $\sum_{i=0}^{n-1}f_{x_i,y}=\sigma$ for all $y$. The advantage of this approach is that the force on the bond at $x_i$ can be solved explicitly as an integral, $f_i=\int_0^Lf_{x_i,y} dy$. We consider the force on $x_0=0$ for simplicity:
\begin{equation}
\begin{aligned}
    f_0 & = \int_{0}^{L}f_{0y}dy \\
    & = \sigma\sum_{q=0}^{n-1}\int_{x_q}^{x_{q+1}}\frac{\cosh\left(\frac{L/2-y}{\ell}\right)}{\sum_{r=0}^q\cosh\left(\frac{L/2-y+x_r}{\ell}\right)+\sum_{r=q+1}^n\cosh\left(\frac{L/2-x_r+y}{\ell}\right)}dy.
\end{aligned}
\end{equation}
 In the second line, the integral is split up into $n$ integrals from one closed bond to the next in order to resolve the absolute value in Eq.~\eqref{eq:force_from_to} i.e., he total force on a closed bond is effectively calculated by integrating the contributions from every gap between two neighbouring closed bonds and then summing over all such pairs. The integrals can be simplified with the substitution $z = e^{y/\ell}$ 
and using $\cosh(x) = \frac{1}{2}\big(e^{x}+e^{-x}\big)$, so that 
\begin{equation}\label{eq:continuous model variable transform}
    f_0 
    =\sigma \ell \sum_{q=0}^{n-1}\int_{e^{x_q/\ell}}^{e^{x_{q+1}/\ell}} \frac{e^{-L/2\ell}z+e^{L/2\ell}z^{-1}}{a_qz^2+b_q} dz
\end{equation}
with 
\begin{equation}\label{eq:definition_bi}
\begin{aligned}
    a_q &= \sum_{r=0}^qe^{-(L/2+x_r)/\ell}+\sum_{r=q+1}^{n-1}e^{(L/2-x_r)/\ell} \\
    b_q &= \sum_{r=0}^qe^{(L/2+x_r)/\ell}+\sum_{r=q+1}^{n-1}e^{-(L/2-x_r)/\ell}.
\end{aligned}
\end{equation}
The integral in Eq. \eqref{eq:continuous model variable transform} can be evaluated, see appendix \ref{app:forceIntegral}, which leads to 
\begin{equation}\label{eq:continuous_model_with_correction}
\begin{aligned}
    f_0  
    =\sigma \ell\sum_{q=0}^{n-1}&\bigg(\frac{1}{2} \Big(\frac{e^{-L/2\ell}}{a_q}-\frac{e^{L/2\ell}}{b_q}\Big) \log\Big(\frac{\sum_{r=0}^{n-1}\cosh\big((L/2-|x_r-x_{q+1}|)/\ell\big)}{\sum_{r=0}^{n-1}\cosh\big((L/2-|x_r-x_q|)/\ell\big)}\Big)\\
    &+\frac{1}{2} \Big(\frac{e^{-L/2\ell}}{a_q}+\frac{e^{L/2\ell}}{b_q}\Big)(x_{q+1}-x_q)/\ell \bigg).
\end{aligned}
\end{equation}
Note that the first term of Eq.~\eqref{eq:continuous_model_with_correction} vanishes if the bond configuration is invariant under exchanging $x_q$ and $x_{q+1}$ since the denominator and numerator within the log will be the same. In particular this is the case when the bond configuration is symmetric around the middle of the gap between $x_q$ and $x_{q+1}$. 
We assume that typical configurations are approximately symmetric and that the first term in each sum is negligible compared to the second. This is because force propagation is left-right symmetric and we do not expect a systematic difference of bond distributions based on the direction.
We thus approximate $f_0$ by plugging in Eq.~\eqref{eq:definition_bi} into the second term of Eq.~\eqref {eq:continuous_model_with_correction}
\begin{equation}\label{eq:force_approximation}
    f_0\approx\sum_{q=0}^{n-1} \frac{\sigma(x_{q+1}-x_q)}{2} \cdot\bigg(\Big(\sum_{r=0}^qe^{-x_r/\ell}+\sum_{r=q+1}^{n-1}e^{(L-x_r)/\ell}\Big)^{-1} + \Big(\sum_{r=0}^qe^{x_r/\ell}+\sum_{r=q+1}^{n-1}e^{-(L-x_r)/\ell}\Big)^{-1}\bigg).
\end{equation}
Each term in the sum corresponds to the force from each gap between two bonds that propagates to the bond at $0$. The force applied on the gap between $x_q$ and $x_{q+1}$ is $\sigma(x_{q+1}-x_q)$, which is divided by two as it propagates to both directions from the gap, which is multiplied by the factor determining how much of the force actually arrive at $0$.  
Figure~\ref{fig:continuous force approximation}  illustrates how $f_0$ in Eq.~\eqref{eq:force_approximation} can be interpreted in this way. The force $\sigma\cdot d$ applied over a gap of length $d$ is split up equally between the bonds to its right and left. It is distributed to the bonds according to the exponential decay with distance from the gap.

\begin{figure}[tb]
    \centering
    \includegraphics[width=0.9\textwidth]{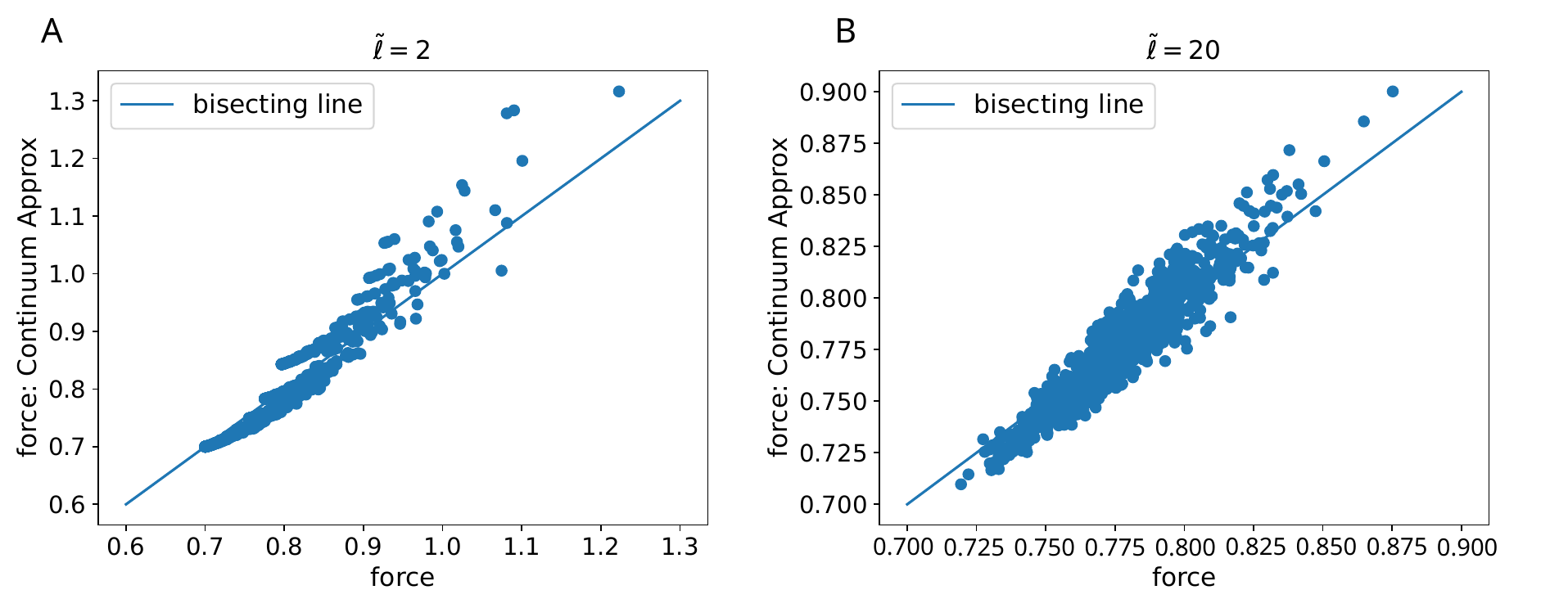}
    \caption[]{Continuum approximation for the force on a closed bond: The force on each bond as given by the continuum approximation, Eq.~\eqref{eq:force_approximation}, is compared  with the corresponding exact force given by Eq.~\eqref{eq:general force} in a system with random configuration of open and closed bonds. If the approximation of force would match with the force perfectly the data points would lie on the bisecting line shown by the blue line. The parameters are: $\tilde{\ell} = \ell/\delta L = 2$ (A), and $\tilde{\ell} = \ell/\delta L = 20$ (B), $N=200$, $K= k_{\text{on}}/(k_{\text{on}}+k_{\text{off},0}) = 0.9$, and $\tilde{\sigma} = \sigma\delta L/f_d =0.7$.  }
    \label{fig:approximation accuracy}
\end{figure}
 To verify the accuracy of this approximation of the general model of force distribution, we compare the force calculated at each closed bond following the full discrete model, Eq.~\eqref{eq:general force}, to the continuum approximation of the force, Eq.~\eqref{eq:force_approximation}, for fixed $K = k_{\text{on}}/(k_{\text{on}}+k_{\text{off},0})$, $N$, the dimensionless external force: $\tilde{\sigma} = \sigma \delta L/f_d$, and the dimensionless decay length scale of the force: $\tilde{\ell} = \ell/\Delta L$, see Fig.~\ref{fig:approximation accuracy}. Forces are non-dimensionalized by rescaling with the characteristic force scale $f_d$. We consider a random configuration of open and closed bonds, where the probability of each bond being open is $K$. We see that the approximation is reasonably accurate but 
over-estimates the force for larger forces. 
 
 We can also verify the validity of the approximation by obtaining the expressions of force on a closed bond in the limit of small and large of the decay length, $\ell$ from Eq.~\eqref{eq:force_approximation}. 
First, in case of large decay length scale or $\ell\rightarrow \infty$ we find,
\begin{equation}
\begin{aligned}
    \lim_{\ell\rightarrow\infty} f_0 &= \sum_{q=0}^{n-1} \frac{\sigma(x_{q+1}-x_q)}{2} \cdot\bigg(\Big(\sum_{r=0}^q1+\sum_{r=q+1}^{n-1}1\Big)^{-1}+ \Big(\sum_{r=0}^q1+\sum_{r=q+1}^{n-1}1\Big)^{-1}\bigg)\\
    &=\sum_{q=0}^{n-1} \frac{\sigma(x_{q+1}-x_q)}{2} \cdot\frac{2}{n} = \frac{\sigma L}{n},
\end{aligned}
\end{equation}
which corresponds to the expression of global load sharing in ref.~\cite{mulla2018crack}.

In the other limit, $\ell\rightarrow 0$, the force decays over a short length. Thus only the forces from the nearest gaps can contribute to the force on the closed bond, meaning force propagated to the closed bond at $0$ is only influenced by the contribution from $q = 1$ and $q = n-1$ of Eq.~\eqref{eq:force_approximation}, which leads to
\begin{equation}\label{eq:force_approx_local}
    \lim_{\ell\rightarrow 0} f_0 
    = \frac{\sigma x_1}{2} + \frac{\sigma(L-x_{n-1})}{2} = \frac{\sigma l_0}{2}. 
\end{equation}
These results match the 
exact results for the global and local load-sharing models, respectively.
This indicates that considering a continuously applied force density over the entire length of the system as opposed to forces applied at each of the $N$ sites to approximate the distribution of forces on the closed bonds in a system with non-local interaction is a reasonable approach for modeling the dynamics of such systems.

We can approximate Eq.~\eqref{eq:force_approximation} further to obtain a simple expression for small, but finite decay length scale $\ell$.
This approximation is non-trivial, because we cannot approximate the force $f_0$ by a Taylor series expansion at $\ell=0$, since all derivatives of $f_0$ at $\ell=0$ are zero. We use a more intuitive and physical approach instead. Instead of considering the force contribution of the closest gaps as in Eq.~\eqref{eq:force_approx_local}, we consider the force contributions from all the gaps, but neglect terms in those contributions that are comparatively small. Specifically, we only retain the largest terms of each gap's force contribution in Eq.~\eqref{eq:force_approximation}, 

\begin{equation}\label{eq:small l approximation 1}
    f_0
    =\sum_{q=0}^{n-1} \frac{\sigma(x_{q+1}-x_q)}{2} \cdot\Big(e^{-(L-x_{q+1})/\ell}+ e^{-x_q/\ell}\Big).
\end{equation}

In order to further simplify the expression, we might only keep the largest of the exponential terms, that is
\begin{equation}\label{eq:small l approximation 2}
    f_0 \approx\frac{\sigma x_1}{2} + \frac{\sigma (L-x_{n-1})}{2} +\frac{\sigma (x_2-x_1)}{2}e^{-x_1/\ell} + \frac{\sigma (x_{n-1}-x_{n-2})}{2}e^{-(L-x_{n-1})/\ell}
\end{equation}
We point out that the first two terms correspond to the limit $\ell\rightarrow 0$. With Eqs.~\eqref{eq:small l approximation 1} and \eqref{eq:small l approximation 2}, we have two distinct approximations of the force on a bond for small $\ell$.

\begin{figure}[tb]
    \centering
    \includegraphics[width=\textwidth]{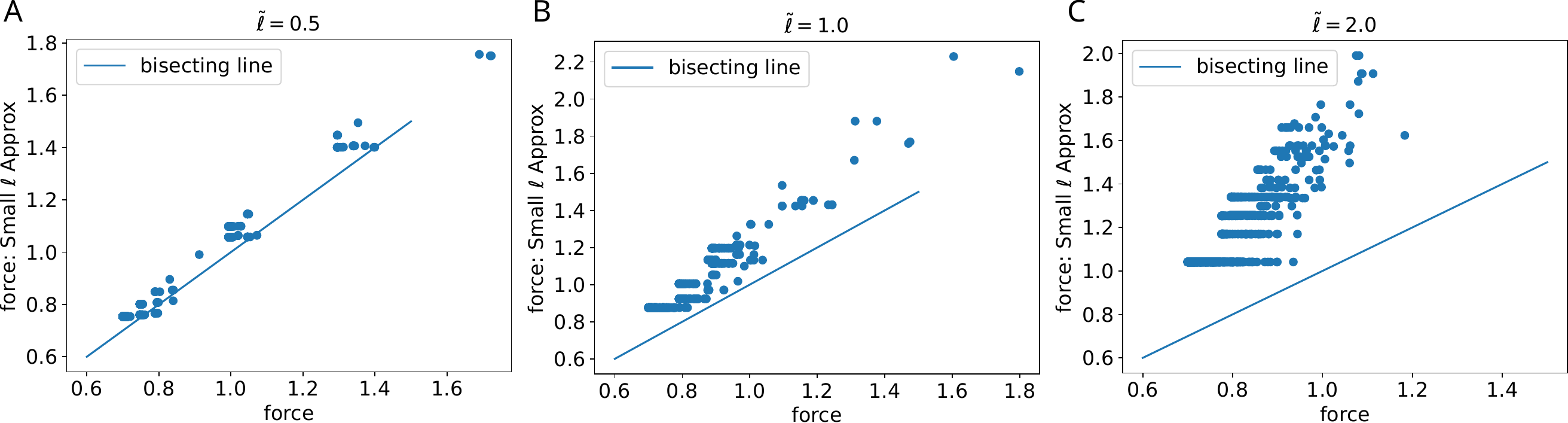}
    \caption[Accuracy of approximation for different decay lengths]{Accuracy of the approximation for small decay lengths: Comparison of the small $\ell$ approximation of the force, Eq.~\eqref{eq:small l approximation 2}, with the force given by the full model, Eq.~\eqref{eq:general force}, for a random configuration of open and closed bonds and three different values of the force distribution length: $\tilde{\ell} = \ell/\delta L = 0.5 $ (A), $\tilde{\ell}= 1 $ (B), and $\tilde{\ell} = 2$ (C). A perfect match of the approximation and the full model means the data would lie on the blue bisecting lines. The other parameters are: $N=200$, $\tilde{\sigma} =\sigma \delta L/f_d = 0.7$, and $K= k_{\text{on}}/(k_{\text{on}}+k_{\text{off},0}) = 0.9$ for all the plots.  
    }
    \label{fig: force v force plots}
\end{figure}
Lastly, we compare the approximation in Eq.~\eqref{eq:small l approximation 2} with the full discrete model given by Eq.~\eqref{eq:general force}. We consider a random configuration of open and closed bonds, where the probability of a bond being open is $K$. We plot the force on each closed bond calculated using the full discrete model, Eq.~\eqref{eq:general force}, on the horizontal axis and the small $\ell$ approximation of force, Eq.~\eqref{eq:small l approximation 2}, on the vertical axis for varying $\ell$ but fixed parameters otherwise. This comparison is shown in Fig.~\ref{fig: force v force plots}. 
For small $\ell$ (Fig.~\ref{fig: force v force plots}A, B) the data points are very close to the diagonal, indicating that the approximation is very good.  However, with increasing $\ell$, the data points tend to move further from the diagonal, in line with our expectation as we derived the approximation explicitly for small $\ell$. Our approximation tends to overestimate the force on bonds for high $\ell$ because of the large $\ell$-dependent terms in Eq.~\eqref{eq:small l approximation 2}. We also note that the spread of the data points is wider for larger $\ell$. 

\section{Bond dynamics and rupture conditions}
\label{sec:Rupture_conditions}

\begin{figure}[tb]
    \centering
    \includegraphics[width=\textwidth]{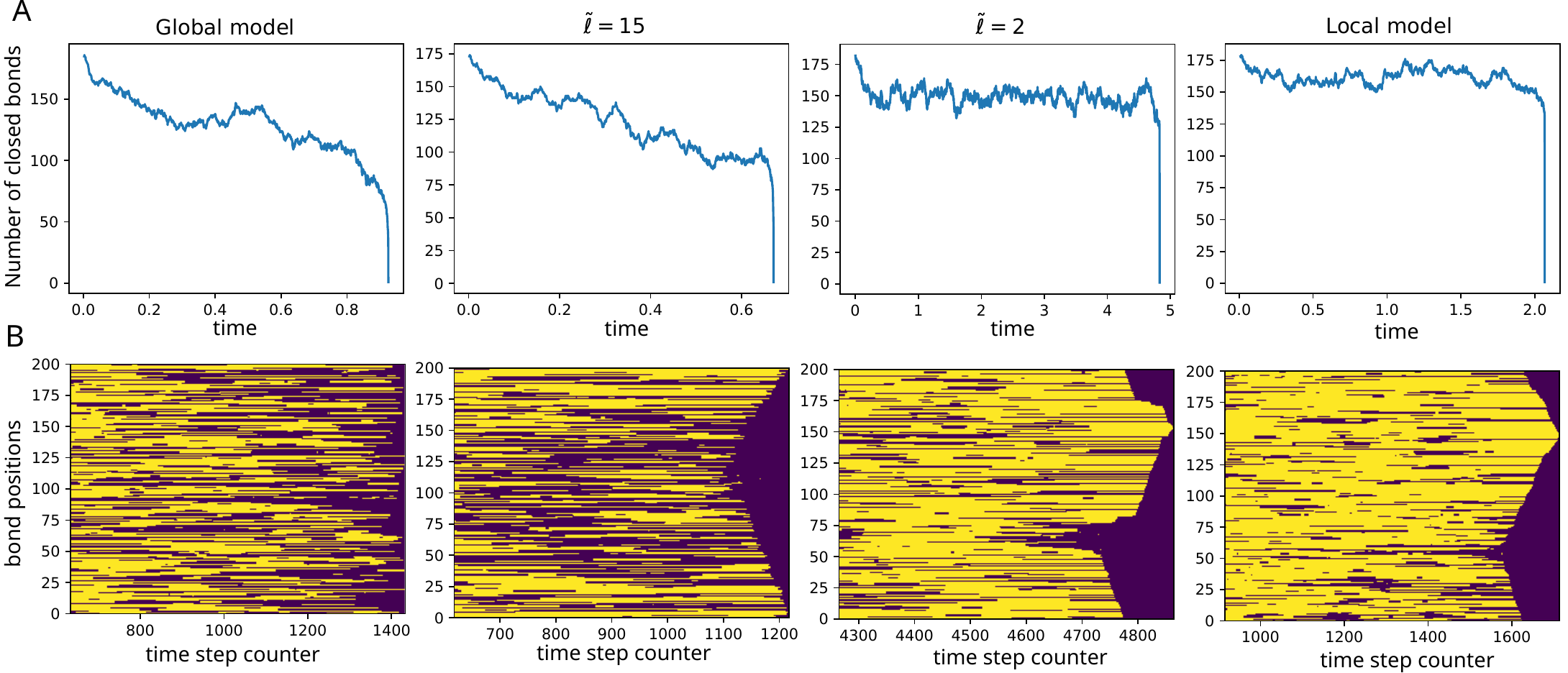}
    \caption[Kymographs]{Time series from simulations of bond rupture: A) The number of closed bonds as function of time for different force localization lengths $\tilde\ell$, left to right: $\tilde\ell\rightarrow\infty$ (global model), $\tilde\ell=15$, $\tilde\ell=2$, 
    and $\tilde\ell\rightarrow 0$ (local model). B) Kymographs indicating the spatial distribution of closed and open bonds (yellow and purple, respectively). The horizontal axis shows the time steps of the Gillespie algorithm, the vertical axis the bond positions. The total number of bonds is $N$ = 200. The force density is chosen to be slightly below the critical force for each case, $\tilde{\sigma} = 1.1$ for $\tilde\ell\rightarrow\infty$, $\tilde{\sigma} = 1.1$ for $\tilde{\ell}=15$, $\tilde{\sigma} = 0.8$ for  $\tilde{\ell}=2$, and $\tilde{\sigma} = 0.6$ for $\tilde\ell \rightarrow 0$ (left to right).}\label{fig:ClosedBonds}
\end{figure}

In the previous sections we introduced a model of bond dynamics with non-local interactions. Here, we present results from simulations of the full model (see Appendix~\ref{sec:Appendix_gilespieal} for details of the simulations), and contextualise them analytically. Figure \ref{fig:ClosedBonds}A shows exemplary time courses of the number of closed bonds from simulations of our model. The plots show  the two limiting cases $\ell \rightarrow \infty$ (global model, left) and  $\ell \rightarrow 0$ (local model, right) as well as cases with finite  $\ell$. 
Note that the force density $\sigma$ is different in the four cases, the values were chosen such that the force density is below the critical value that we will discuss below. The plots  
show in all cases that the system resides in a metastable state for some time with some bonds opening and closing. Eventually, all bonds open up in quick succession until none are left as observed in previous works \cite{bell1978models,erdmann2004stability,mulla2018crack}. We call this event the rupture of the system and the first time at which all bonds are open the rupture time. 
In Fig.~\ref{fig:ClosedBonds}B, we show the bond dynamics of the same simulations in a kymograph (yellow indicates a closed bond, purple an open bond). Here the time axis shows the steps of the Gillespie simulation, note that these have random duration and are shorter when events occur with high rate. These kymographs thus enlarge the the quick succession of bond openings during the rupture event. In the limit of the global model, the spatial distribution of closed bonds is thinning out more or less uniformly, whereas for finite localization length $\ell$ and for the local limit, rupture is a more localized event: A crack, a region without closed bonds, appears and  starts spreading through the entire system. This phenomenon, referred to as crack initiation, has been described previously in the limit $\ell\rightarrow 0$ \cite{mulla2018crack}. Clear examples are visible for $\ell=2$ and for the local model. Open bonds within the crack rarely close again, but remain open for the rest of the simulation, while bond reformation is more common for larger $\ell$. Eventually, the spreading of the crack leads to rupture. 
This behavior is seen in the local limit $\ell\rightarrow 0$ and for small, nonzero  $\ell$, while spatially uniform opening is seen in the global limit. For intermediate values of  $\ell$ (e.g. $\tilde{\ell} =15$), we observe that bonds open uniformly, while a crack forms simultaneously, indicating that these two pathways to rupture coexist for intermediate $\ell$, while one of them becomes dominant in the limiting cases. 

\subsection{Rupture conditions in the limits $\ell\rightarrow 0$ and $\ell\rightarrow\infty$}
\label{sec:Rupture_global_local}

Before considering the general case, we discuss conditions for rupture in the  limiting cases of global and local coupling, specifically  a critical force and a critical gap size, respectively.
In the global case, the on and off rates are independent of the bond position. Therefore the total number of closed bonds $n$ is enough to characterize a steady state. The number of closed bonds reaches a steady state when the total rates of closing and opening bonds are equal,
\begin{equation}\label{eq:critical bond number global}
     (N-n)\cdot k_{\text{on}} = n\cdot k_{\text{off}}= n\cdot k_{\text{off},0}\cdot \exp\Big(\frac{N\tilde{\sigma} }{n}\Big),
\end{equation} 
For sufficiently small $\tilde\sigma$, this equation has two solutions, which can only be determined numerically, for large $\tilde{\sigma}$  none. The condition for solutions to exist is, 
\begin{equation}\label{eq:sigma - K critical}
  K=\frac{k_{\text{on}}}{k_{\text{on}}+k_{\text{off},0}}  \geq 
  \frac{\tilde{\sigma}}{e^{-(1+\tilde{\sigma})}+\tilde{\sigma}}.
\end{equation}
Equality in the expression defines a critical value, either a critical force density $\tilde\sigma_c$ (for given $K$) or a critical binding strength $K_c=\frac{\tilde{\sigma}}{e^{-(1+\tilde{\sigma})}+\tilde{\sigma}}$ (for given $\tilde\sigma$). This result corresponds to the one derived by Bell \cite{bell1978models}. For completeness its derivation is included in 
Appendix \ref{sec:details glocal model}. This condition is also related to the criterion that distinguishes the two regimes of rupture time in the global load-sharing model with dynamic loading studied in previous work \cite{seifert_rupture_2000}. 
Beyond the critical force, rupture of the system is inevitable in the deterministic version of the model \cite{bell1978models}. This is also seen in the stochastic model, but here the system will still rupture if the number of closed bonds drops below the steady-state value due to stochastic fluctuations, even for $K>K_c$ or $\tilde\sigma<\tilde\sigma_c$, as studied previously \cite{erdmann2004adhesion} and shown in Fig.~\ref{fig:ClosedBonds} (rightmost case). 

In turn, considering the limit of local force distribution, we observe that the rupture is caused by a sufficient number of consecutive open bonds, i.e., a crack or gap. To obtain a steady state condition, we equate the on-rates of all the $d/\delta L -1$ open bonds within one gap of length $d$ and the off-rates of the two neighbouring closed bonds. This will define a criticality condition for rupture in terms of a critical gap size $d_c$. We define the dimensionless quantities $\tilde{d} = d/\delta L$ to refer to the number of sites within the distance $d$ and likewise $\tilde{d}_c = d_c/\delta L$ and obtain the steady state condition
\begin{equation}\label{eq:condition d critical}
    (\tilde{d}_c-1)\cdot k_\text{on} = k_{\text{off},0}\cdot\exp\left(\frac{\tilde{\sigma}}{2} \left(\tilde{d}_c+\frac{1}{K}\right)\right).
\end{equation}
In this expression, we treat all other gaps in a mean-field like manner and assume that they have the same size, setting the distance to the next closed bond on the other side to $1/K$, the average distance in a system without force. From this condition, solving for $\tilde{d}_c$, we obtain a critical gap size
\begin{equation}\label{eq:critical d local}
    \tilde{d}_c = 1-\frac{2}{\tilde{\sigma}}\cdot W_{-1}\bigg(-\frac{\tilde{\sigma} K}{1-K}\exp\big(\frac{\tilde{\sigma}}{2}(1/K+1)\big)\bigg),
\end{equation}
see Appendix \ref{sec:details glocal model}. 
Here $W_{-1}$ is a branch of the Lambert $W$ function, that is the inverse function of $x\mapsto x\cdot e^x$. When the gap between two closed bonds is larger than this critical value, the gap will grow and rupture will be initiated. The estimated critical gap size agrees with the result in ref.~\cite{mulla2018crack}, with an additional incorporation of the force contributions of nearest-neighbor gaps. 

\begin{figure}[tb]
    \centering
    \includegraphics[width=0.8\textwidth]{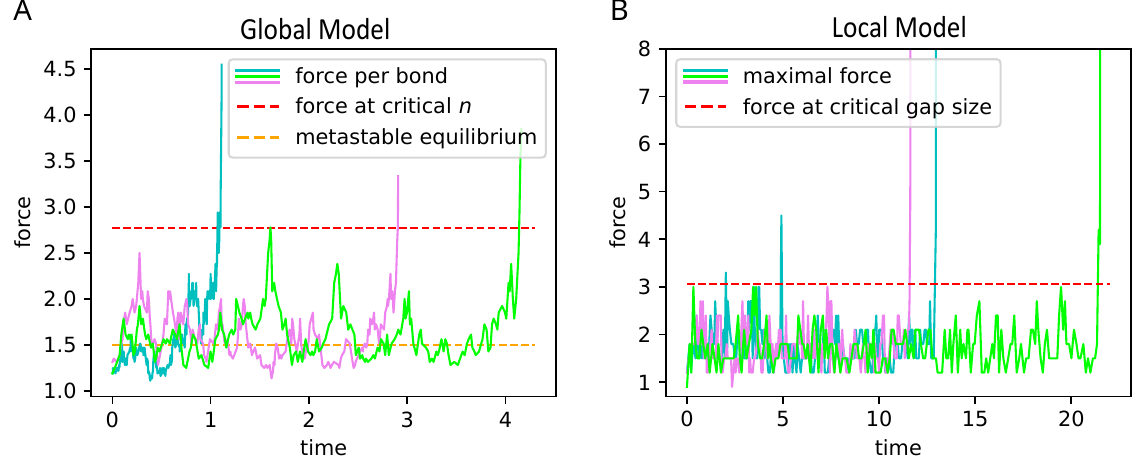}
    \caption[]{Forces and rupture conditions in the local and global limit: A) Global coupling: Time series of the force per bond for three simulations. The lines indicate the force per bond in the metastable state(orange line) and the  critical force (red line). These were calculated numerically via the two critical values of $n$ according to Eq.~\eqref{eq:critical bond number global}. The parameters are $\tilde{\sigma} =1$, . B) Maximal force (at the largest gap) over time for three simulations with $\tilde{\sigma} =0.6$, and the force at the critical gap size according to Eq.~\eqref{eq:critical d local}. $N=50$ (left) and $N=100$ (right) and $K=0.9$.}\label{fig:criticality criteria under test}
\end{figure}

In Fig.~\ref{fig:criticality criteria under test}, we check the rupture conditions with simulations. Figure~\ref{fig:criticality criteria under test}A shows several time series of the non-dimensional force per bond as obtained from simulations of the global force distribution model. The red and orange dashed line indicate the forces per bond corresponding to the two solutions of Eq.~\eqref{eq:critical bond number global}. The force time series fluctuates around the (meta-)stable solution (orange line) and rapid rupture is coupled to the force crossing the red line, which marks the deterministically unstable solution.

In the local force distribution case, we plot the non-dimensional maximal force on a bond at each time step, which typically occurs at the largest gap that eventually becomes a critical crack and initiates rupture (Fig.~\ref{fig:criticality criteria under test}B). This maximal force also fluctuates around a stable value, but increases rapidly once the force exceeds the value corresponding to the critical gap size. This force is calculated based on Eq.~\eqref{eq:critical d local} using the local model and shown as the red dashed line. However, occasionally the system is saved from rupture even beyond our estimated critical point. Such events are more pronounced than in the local model, likely because the critical gap size is an approximation describing neighboring gaps in a mean-field-like manner. By contrast, the criticality condition in the global case is the exact stability condition in the deterministic limit. 

\subsection{Rupture condition for small $\ell$}
\label{sec:rupture_cond}

 \begin{figure}
    \centering
    \includegraphics[width=0.4\textwidth]{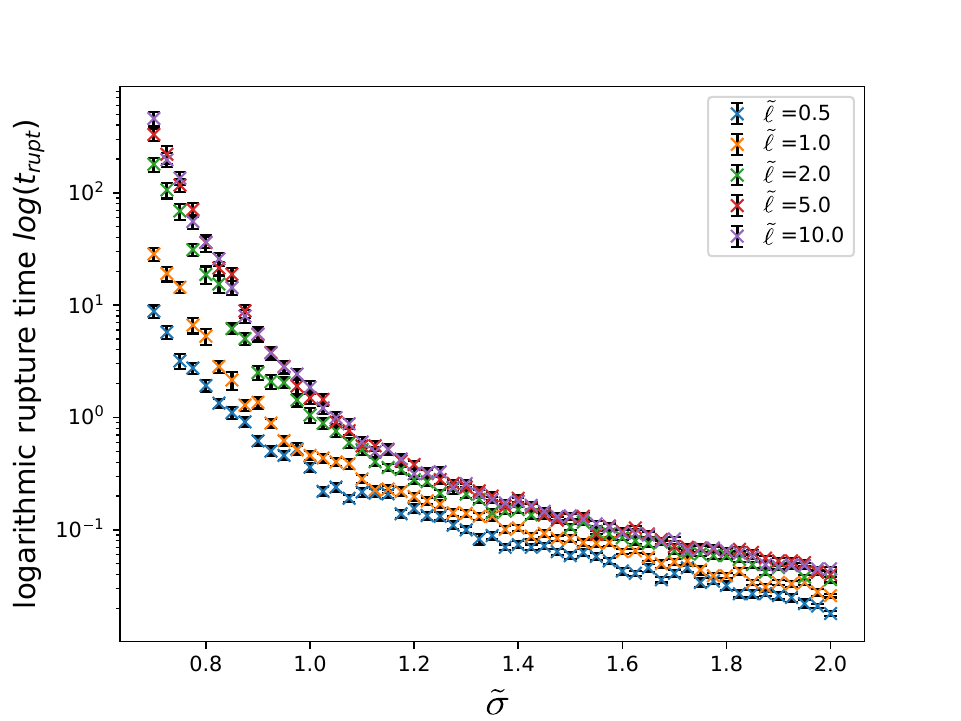}
    \caption[Average rupture time as a function of the external force]{Average rupture time as a function of the external force ($\tilde{\sigma}$) for varying decay length $\ell$: The average for one data point is taken over 50 trials. The curves look like staggered versions of one another. Generally, higher $\ell$ leads to the load being more equally distributed among the closed bonds and therefore a higher rupture time. All curves display the two distinct regimes of metastable and unstable behaviour. $N=20$, $K=0.9$.}
    \label{fig:generalized rupture time}
\end{figure}

To understand how a general force distribution length scale $\ell$ can influence the stability of an ensemble of bonds, we first quantified the rupture time of our general model for varying external force $\tilde{\sigma}$. The rupture time decreases with increasing external force, similar to the rupture times of the limiting cases shown in ref.~\cite{mulla2018crack}. Furthermore, as shown in Fig.~\ref{fig:generalized rupture time}, we observed that the average rupture time at any value of the force $\tilde{\sigma}$ increases with increasing decay length $\tilde{\ell}$. However, the increase is seen to saturate as the rupture time data points for $\tilde{\ell} = 2.0$, $\tilde{\ell} = 5.0$ and $\tilde{\ell} = 10.0$ overlap, indicating that the rupture time is strongly influenced by $\tilde{\ell}$ only for $\tilde{\ell} \lesssim 2.0$.

In Sec.~\ref{sec:continuum_approx}, we have found two approximations for the force on one closed bond that apply to this regime of small $\ell$: an intuitive and rather accurate continuum approximation given by Eq.~\eqref{eq:force_approximation} and an even simpler expression for small $\ell$ given by Eq.~\eqref{eq:small l approximation 2}. In this section, we use these approximations to determine a rupture criterion for the case of small, but finite $\ell$.

Similar to the rupture condition of the local model shown in Sec.~\ref{sec:Rupture_global_local}, for small $\ell$, we expect the largest gap size to be the decisive factor for rupture. 
However, in this limit, the system is not entirely determined by the largest gap size, but the bonds in its neighbourhood also play a role. We treat the gaps between these bonds with the same mean field approach as for the local model above. That is, we consider a configuration  where all the closed bonds are equidistantly distributed  except for one larger gap of size $d$ between $x_{n-1}$ and $x_0$, that is $x_i = i\delta L/K$ with $(n-1)/K=(L-d)/\delta L$. We then plug this expression into the approximation of force for small $\ell$ given by Eq.~\eqref{eq:small l approximation 2}. This gives the force at $i=0$, at the boundary of the gap, as 
\begin{equation}\label{eq:critical_gap_size_small_l}
    f_0  \approx\frac{\sigma d}{2} +\frac{\sigma }{2K}\Big(1 +e^{-\delta L/K\ell} + e^{-d/\ell}\Big).
\end{equation}
The force at $i=n-1$, the closed bond at the other boundary of the gap is the same. The critical gap size is then again determined from the steady state of the number of closed bonds, where the rate of an open bond closing inside the largest gap is the same as the rate of the closed bonds at the boundary of the gap opening. The steady state of the system can be defined by
\begin{equation}\label{eq:CritGapCondition}
    \tilde{d}_c k_{\text{on}} = 2k_{\text{off,0}}\exp(\frac{f_0}{f_d}) 
\end{equation}
with $f_0$ from Eq.~\eqref{eq:critical_gap_size_small_l}. From this condition, we obtain an implicit equation for the critical gap size
\begin{equation}\label{eq:CritGapl1}
    \tilde{d}_c\cdot\Big(\frac{1}{K}-1\Big) = \exp\Big(\frac{\tilde{\sigma} \tilde{d}_c}{2} +\frac{\tilde{\sigma }}{2K}\big(1 +e^{-1/K\tilde{\ell}} + e^{-\tilde{d}_{c}/\tilde{\ell}}\big)\Big).
\end{equation}

\begin{figure}[tb]
    \centering
    \includegraphics[width=\textwidth]{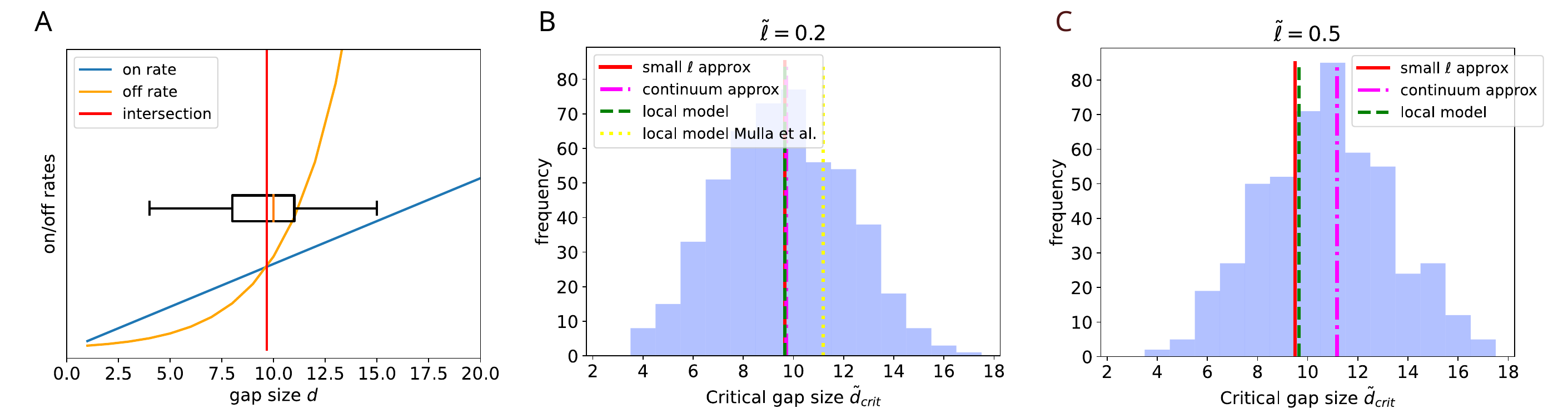}
    \caption[Critical gap size for small $l$]{Exemplary visualisation of the critical gap size: A) The plot displays the on-rate and approximated off-rate as a function of the gap size $d$ for fixed parameters $N,K,\sigma, \ell$. 
    Intersections of on- and off-rate mark the critical gap size - emphasized by the red line. The critical gap size has also been inferred from simulations. The results are included as a boxplot based on 500 simulations for $\tilde{\ell}=0.2$. B) and C) Histogram of critical gap size inferred from simulations, 
    for $\tilde{\ell}=0.2$ (B) or $\tilde{\ell}=0.5$ (C) respectively including numerical estimates based on Eq.~\eqref{eq:CritGapl1} (red line), Eq.~\eqref{eq:approximation force specific} (magenta dash-dotted line), Eq.~\eqref{eq:critical d local} (green dashed line) and approximation from Mulla et al. \cite{mulla2018crack}. See appendix \ref{sec:relative error critical gap size} for the exact values. Parameters: $N=200, K=0.9, \tilde{\sigma}=0.7$ N =200, K=0.9, $\tilde{\sigma} =0.7$, $\tilde{\ell} =0.2$.}
    \label{fig:critical gap size small l}
\end{figure}

Fig.~\ref{fig:critical gap size small l}A illustrates how the critical gap size can be determined graphically by the intersection of left- and right-hand side of  Eq.~\eqref{eq:CritGapl1}, which correspond to the on and off rates. The computed critical gap size $\tilde{d}_{c}$ (here $\simeq 9.6$) is marked with a vertical red line.

Furthermore, we also compare our estimate of the critical gap size with simulations. We obtain a numerical estimate of the critical gap size from the simulations by extracting the maximal gap size in all bond configurations excluding the final $N$ time steps. The end of the simulation is marked by the opening of the last closed bond. Since $N$ is the system size the rupture will take no more than $N$ time steps after its initiation. Hence by excluding the last $N$ simulation steps, we exclude the large unstable gaps that occur during the rupture event. In previous works, the critical gap size has occasionally been investigated with ablation experiments \cite{mulla2018crack, mulla2018crosslinker}, where, after an equilibration period, a gap of a certain size is introduced and the fraction of simulations that rupture within a fixed time is measured. 
Although computationally more expensive than the previous ablation method \cite{mulla2018crack, mulla2018crosslinker} especially for large $N$, our method mitigates the influence of simulation run times.

The critical gap size $\tilde{d}_c$ is $9.6\pm2.5$ for small $\tilde{\ell} = 0.2$; The distribution of critical gap sizes in simulation are shown by the boxplot in Fig.~\ref{fig:critical gap size small l}A and by a histogram in Fig.~\ref{fig:critical gap size small l}B. We compare the estimation of critical size from Eq.\eqref{eq:CritGapl1} (red line) with the estimation of critical gap size of a local model Eq.~\eqref{eq:critical d local} (green dashed line) and the critical gap size of a local model without the additional effect of neighbouring bonds in Mulla et al.~\cite{mulla2018crack} (yellow dotted line). The estimate from Mulla et al.~\cite{mulla2018crack} ($\simeq 11.2$) is larger than the mean critical gap size observed in simulation. However, the critical gap size of a local model by Eq.~\eqref{eq:critical d local}, the estimate by Eq.~\eqref{eq:CritGapl1}, and the one obtained from simulations are in good accordance, see Fig.~\ref{fig:critical gap size small l}B. 

Although we could accurately estimate the critical gap size when the force distribution lengthscale is very small, i.e. $\tilde{\ell} = 0.2$, using the approximation of force for small $\ell$ given by Eq.~\eqref{eq:small l approximation 2} or by considering a local model, we can improve the accuracy of our estimate of the critical gap size especially for larger $\ell$ by using the more accurate approximation for the force from Eq.~\eqref{eq:force_approximation}, 
instead of that of Eq.~\eqref{eq:small l approximation 2}. With the same mean field treatment of the bonds outside the large gap as above, the force on the two bonds at the boundary of the large gap is

\begin{equation}\label{eq:approximation force general}
\begin{aligned}
     f_0\approx&\frac{\sigma \delta L}{2K} \sum_{q=0}^{n-2} \cdot\bigg(\Big(\sum_{r=0}^qe^{-r/K\tilde{\ell}}+\sum_{r=q+1}^{n-1}e^{(N-r/K)/\tilde{\ell}}\Big)^{-1}+ \Big(\sum_{r=0}^qe^{r/K\tilde{\ell}}+\sum_{r=q+1}^{n-1}e^{-(N-r/K)/\tilde{\ell}}\Big)^{-1}\bigg)\\
    &+\frac{\sigma \delta L \tilde{d}}{2} \bigg(\Big(\sum_{r=0}^{n-1}e^{-r/K\tilde{\ell}}\Big)^{-1}+ \Big(\sum_{r=0}^{n-1}e^{r/K\tilde{\ell}}\Big)^{-1}\bigg).
\end{aligned} 
\end{equation}
Here $\tilde{d} = d/\delta L$ is again the dimensionless size of the large gap.
Performing the summation over r, we obtain 
\begin{equation}\label{eq:approximation force specific}
\begin{aligned}
    f_0\approx&\frac{\sigma \delta L}{2K} \sum_{q=0}^{n-2} \bigg((1-e^{-1/K\tilde{\ell}})\Big(1-e^{-(q+1)/K\tilde{\ell}}+e^{(N-(q+1)/K)/\tilde{\ell}}-e^{\tilde{d}/\tilde{\ell}}\Big)^{-1}\\
    &\quad\quad\quad+(1-e^{1/K\tilde{\ell}})\Big(1-e^{(q+1)/K\tilde{\ell}}+e^{-(N-(q+1)/K)/\tilde{\ell}}-e^{-\tilde{d}/\tilde{\ell}}\Big)^{-1}\bigg)\\
    &+\frac{\sigma \delta L \tilde{d}}{2} \bigg(\frac{1-e^{-1/K\tilde{\ell}}}{1-e^{-(N-\tilde{d})/\tilde{\ell}}}+ \frac{1-e^{1/K\tilde{\ell}}}{1-e^{(N-\tilde{d})/\tilde{\ell}}}\bigg),
\end{aligned}
\end{equation}
see Appendix \ref{sec:details calculations force at gap} for details.

For large $N$, we can  approximate the last term as
\begin{equation}
\begin{aligned}
     \frac{\sigma \delta L \tilde{d}}{2} \bigg(\frac{1-e^{-1/K\tilde{\ell}}}{1-e^{-(N-\tilde{d})/\tilde{\ell}}}+ \frac{1-e^{1/K\tilde{\ell}}}{1-e^{(N-\tilde{d})/\tilde{\ell}}}\bigg) \approx \frac{\sigma \delta L\tilde{d}}{2} (1-e^{-1/K\tilde{\ell}}).
\end{aligned}
\end{equation}
Finally, we compute the force in the approximation \eqref{eq:approximation force specific} numerically, and equating again on- and off-rates using this approximation for $f_0$, we receive another estimate of the critical gap size.

In Fig.~\ref{fig:critical gap size small l} B and C, we also compare the observations from simulation with the critical gap size estimated based on the force on a closed bond at the boundary of a gap approximated by Eq.~\eqref{eq:approximation force specific} (magenta dash-dotted line).
 We observe that for $\tilde{\ell} = 0.2$ (Fig.~\ref{fig:critical gap size small l} B) using the more accurate approximation of the force does not noticeable improve the estimate,
indicating that for $\tilde{\ell} = 0.2$ the general model behaves like the local model, which agrees with our previous observations. However, for $\tilde{\ell} = 0.5$ (Fig.~\ref{fig:critical gap size small l} C), we observe that the estimate of the critical gap size obtained based on Eq.~\eqref{eq:approximation force specific} agrees with the simulation much better than the estimate by Eq.~\eqref{eq:CritGapl1} and the local model. More specifically we observe that the average critical gap size is larger than that of the local model and the estimate considering small $\ell$, showing that a distribution of the force over a larger length $\ell$ stabilizes the system. The results shown here  indicate that our estimates of bond dynamics based on the approximation of the force on a bond in our model with non-local force distribution can explain and interpret the dynamics of such general systems for a broad range of parameters.

\subsection{Bounds on the critical gap size}
\begin{figure}[tb]
    \centering
    \includegraphics[width=0.95\textwidth]{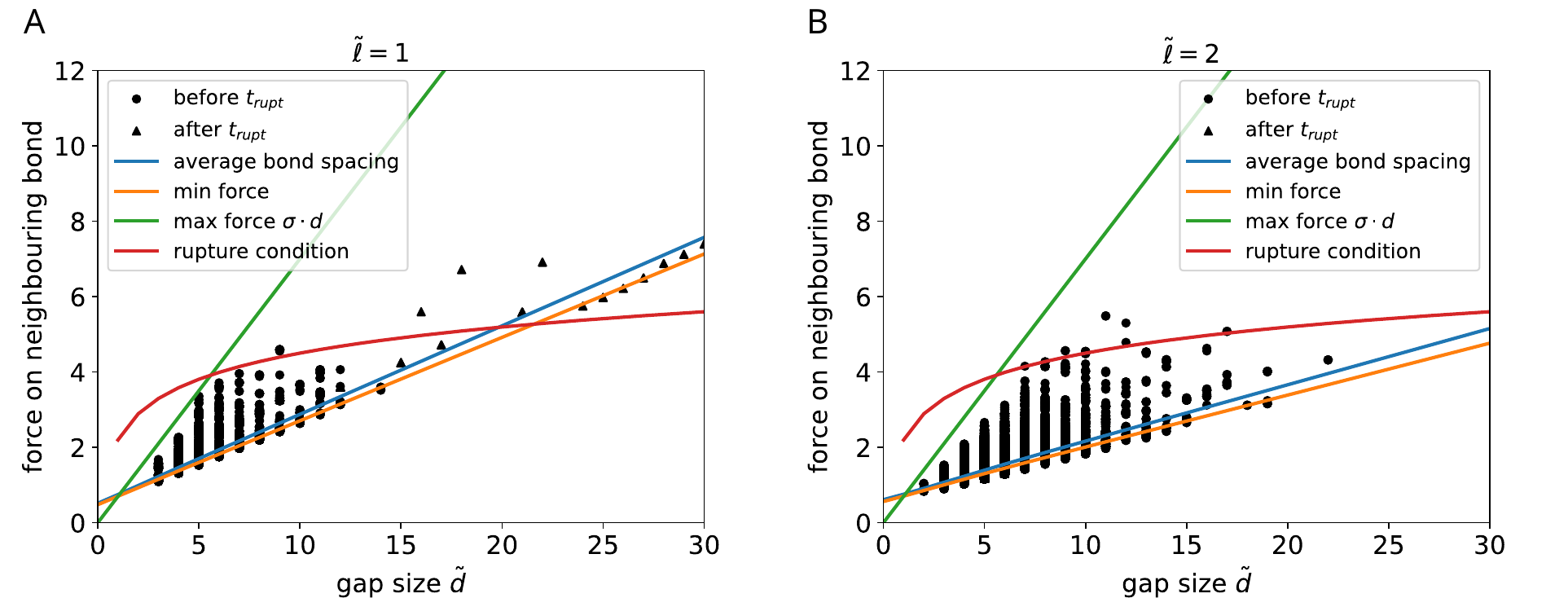}
    \caption[Force at the largest gap over time and sorted by gap size]{
    Plot of the non-dimensional force on the bonds next to the largest gap against the largest gap size quantified from an example simulation both before (circles) and after (triangles) the initiation of rupture defined as $t_{\text{rupt}}$. Analytical estimate of force on bonds next to a large gap: assuming equal spacing of $K^{-1}$ between every other neighbouring bonds (blue line) given by Eq.~\eqref{eq:approximation force specific}, minimum force on bonds next to the gap $f_{0,\text{min}}$ given by Eq.~\eqref{eq:force_approximation_gap_min} (orange line), maximum force on bonds next to the gap $f_{0,\text{max}}$ given by Eq.~\eqref{eq:force_approximation_gap_min} (green line). The condition of rupture is given by Eq.~\eqref{eq:extrema condition} (red line), intersections of the curves with the condition denote the estimate of maximal and minimal gap size at rupture.
        Left: $\tilde{\ell} = 1$, right: $\tilde{\ell} = 2$. $N=200, K=0.9, \tilde{\sigma}=0.7$.}
    \label{fig:comparison force}
\end{figure}

As mentioned above, rupture is not only dependent on the size of the largest gap but also on the neighbouring bonds. Until now, we derived the critical gap size based on a mean field approximation of the neighbouring bond positions, assuming that all gaps except the largest have a characteristic average size. The actual critical gap size will vary around that value, depending on the actual configuration of the other bonds. In this section, we provide upper and lower bounds on the critical gap size, varying the configurations of the other gaps and the corresponding closed bonds. More precisely, we state equations which can numerically determine the largest and smallest gap size at which rupture can be initiated for a given parameter set. This will be done by studying the most favourable and unfavourable bond configuration with respect to the force  distribution.

We can make the following statement regarding the critical gap size in the general model: For a given size $\tilde{d}$ of the largest gap, again located between $x_{n-1}$ and $x_0$, the least force is exerted on the closed bonds at its boundaries when all bonds outside the gap are closed. The force on the boundary bond at $i=0$ in such case can be computed by setting $K=1$ in Eq.~\eqref{eq:approximation force specific}, 
\begin{equation}\label{eq:force_approximation_gap_min}
\begin{aligned}
    f_{0,\min}\approx&\frac{\sigma \delta L}{2} \sum_{q=0}^{n-2} \bigg((1-e^{-1/\tilde{\ell}})\Big(1-e^{-(q+1)/\tilde{\ell}}+e^{(N-(q+1))/\tilde{\ell}}-e^{\tilde{d}/\tilde{ell}}\Big)^{-1}\\
    &\quad\quad\quad+(1-e^{1/\tilde{\ell}})\Big(1-e^{(q+1)/\tilde{\ell}}+e^{-(N-(q+1))/\tilde{\ell}}-e^{-\tilde{d}/\tilde{\ell}}\Big)^{-1}\bigg)\\
    &+\frac{\sigma \delta L \tilde{d}}{2} \bigg(\frac{1-e^{-1/\tilde{\ell}}}{1-e^{-(N-\tilde{d})/\tilde{\ell}}}+ \frac{1-e^{1/\tilde{\ell}}}{1-e^{(N-\tilde{d})/\tilde{\ell}}}\bigg).
\end{aligned}
\end{equation}
On the other hand, the force on the bonds enclosing the largest gap with size $\tilde d$ is maximized if all the other gaps are of the same size $\tilde{d}$. By symmetry, every closed bond will then carry the same force in this case, namely
\begin{equation}\label{eq:force_approximation_gap_max}
    f_{0,\max}=\sigma \delta L\cdot \tilde{d}.
\end{equation}
These forces $f_{0,\min}$ and $f_{0,\max}$ can be used to solve for critical gap sizes in these extreme cases, by plugging them into the steady state condition that equating on and off rates
\begin{equation}\label{eq:extrema condition}
    \tilde{d}_{c,\max/\min}\cdot k_\text{on} = 2k_{\text{off},0}\exp(\frac{f_{0,\min/\max}}{f_d})
\end{equation}
The maximal force leads to the smallest critical gap size $\tilde{d}_{c,\min}$ and vice versa.

To check these bounds, we determined the force at the boundary of the largest gap in simulations. This force is plotted in Fig.~\ref{fig:comparison force} as a function of the size of the largest gap, $\tilde d$ for two values of the force distribution length ($\tilde{\ell} = 1$ and $\tilde{\ell} = 2$).  
To obtain the data, we track the largest gap $\tilde{d}$ at each time step and the largest force on a bond around the largest gap. 
We distinguish data from before rupture time $t_{\text{rupt}}$ and after the rupture time $t_{\text{rupt}}$ by the symbols (black circles and triangles, respectively), the rupture time $t_{\text{rupt}}$ is determined as described in Sec.~\ref{sec:rupture_cond}.
As expected the data shows considerable scatter of the measured force for a given largest gap size. This scatter is bounded by the two bounds estimated above, which we indicate in the plot by the green and orange lines. We note that the mean field approximation of the force, as given by 
Eq.~\eqref{eq:approximation force specific} (show as blue lines) is close to the lower bound and thus typically underestimates the force on the boundaries of the gap.

We also included the force that will initiate rupture for a given gap size determined by the steady state condition in Eq.~\eqref{eq:extrema condition}, shown as the red lines in Fig.~\ref{fig:comparison force}. The intersections of this condition with the estimated force bounds can be used to determine the maximum, minimum or average critical gap size and the corresponding forces. Thus, the region bound by the maximal force (green line), the minimal force (orange line) and the rupture condition (red line) defines a stable region where the ensemble of bond series should not rupture. When comparing with results obtained from simulation with two different $\tilde{\ell}$, we observe that indeed the data points for gap size and force at the boundary, taken before the rupture time $t_{\text{rupt}}$ (circles) mostly fall within this region, while data points taken after $t_{\text{rupt}}$ (triangles) are outside this estimated stable region (only observed for $\tilde{\ell} =1$, as the system did not rupture within the simulation time for $\tilde{\ell} =2$). 
We note that due to the stochastic nature of the simulation, some data before or after rupture initiation fall  outside or inside of the estimated stable region, respectively.

Taken together, these observations show that the mean field calculation gives a good description of stability and rupture of the ensemble of bonds, but our results also indicate its limitations by demonstrating how the relevant forces on the gap boundaries are underestimated.


\section{Discussion and concluding remarks}
We introduced a model of transient bond clusters that incorporates load distribution among the bonds following a given decay length-scale $\ell$ to model systems with force sharing in a neighborhood characterized by that length scale. The limiting cases of this general load distribution model are equal load distribution among all closed bonds (global model) and load distribution among the nearest closed bonds (local model) as studied before \cite{mulla2018crack}.  
Since the full general model is very complex and not easily accessible to intuitive interpretations or analytical calculations, we simplified the full model considering a mean field approximation. 
We investigated the bond dynamics and rupture condition first in the limiting case of equal load distribution and identified the criterion of applied external force in terms of bond kinetics that separates two different regimes of rupture force which is similar to the previous description of different rupture regimes of transient bond clusters \cite{seifert_rupture_2000}. For the other limiting case of load distribution among only the nearest closed bonds, we could also identify a critical gap size that leads to rupture which is a modification of the prediction of \cite{mulla2018crack} by incorporating the neighbouring gaps. We emphasize that in this case, where the force is not the same on all bonds, rupture of the bond cluster is induced by a crack in the system, one large gap in the spatial arrangement of closed bonds (in general, i.e., for intermediate force distribution length, the two rupture pathways, spatially uniform bond opening and crack spreading, coexist). Using the approximation of the general model for non-local/non-global load distribution, we predicted the critical gap size as a function of $\ell$, when the decay length scale, $\ell$, is small but non-zero. 
The analytical predictions show that the critical gap size increases with $\ell$, indicating that the neighboring bonds can stabilize the bond cluster against rupture by crack initiation. As we explicitly consider all the other bond configuration along with the largest crack to define when rupture will be initiated, we were able to identify the bounds of critical gap size for different $\ell$s. The analytical bounds were found to agree well with simulations. 
Our model of transient bond clusters generalizes established models of transient bond clusters\cite{erdmann2004adhesion, erdmann2004stability,seifert_dynamic_2002,seifert_rupture_2000} and can describe the mechanical responses of a wide range of systems where force is not only distributed either among all bonds or among only the nearest bonds \cite{mulla2018crack}, but also everything in between. In this work we focused on a semi-analytical approach, which provides a good approximation for $\tilde\ell \lesssim 2$. An analytical approach to the transition to the global model remains for future research.
One important application we expect for our model is to bonds between cytoskeletal filaments and networks of such filaments. We expect that modeling such systems, in particular with transient crosslinkers, as 
transient bond clusters with non-local interactions 
ca provide a more accurate description than previous work ~\cite{broedersz_cross-link-governed_2010,mulla2018crack,mulla2018crosslinker}, because the model incorporates both the force-dependent unbinding kinetics of individual bonds as well as the mechanically crucial inherent length scale of a semi-flexible filament network determined by the density of filaments, the filaments' persistence length, and the density of crosslinkers~\cite{broedersz2011criticality, alvarado2017force,morse_viscoelasticity_1998,broedersz_modeling_2014,shin_relating_2004}. Moreover, our simplification using a mean field approximation of this general model is useful for drawing analytical insights about the mechanical response, as demonstrated here by the estimate of a rupture condition for the general model.


\newpage


\section*{Conflicts of interest}
There are no conflicts to declare.

\section*{Acknowledgements}
The authors acknowledge funding by the Deutsche Forschungsgemeinschaft (DFG, German Research Foundation) – Project-ID 449750155 – RTG 2756, Project A3 (to S.K.)



%
%

\appendix

\section{Simulations with the Gillespie algorithm}\label{sec:Appendix_gilespieal}

We simulated all variants of the bond dynamics 
as a stochastic simulation implemented with the Gillespie algorithm \cite{gillespie1977exact}. Given that there are $N$ bonds, we represent the state of the system with a vector $\mathbb{S}=(S_{0},\dots,S_{2N-1})$ that contains the number of closed and open bonds at each bond position $x_i$. The even positions $S_{2i}$ in the vector represent the the closed bonds and the odd positions $S_{2i+1}$ refer to the open bonds at the position $i$. Furthermore, we let $\mathbb{X}=(X_{0},\dots,X_{2N-1})$ be the vector of quantities of each type of bond. The possible reactions will have to incorporate the constraints that these quantities can only be $0$ or $1$ and that $X_{2i}+X_{2i+1}= 1$ for all $i\in \{0,\dots,N-1\}$ as a particular bond can only be open or closed exclusively. The initial states of the bonds are chosen independently and are closed with probability $K = k_{\text{on}}/(k_{\text{on}}+k_{\text{off}})$ such as though the force is switched on at the starting time.

There is a total of $2N$ reactions given by the switches of each of the bonds from closed to open or vice versa. The reaction equations are
\begin{alignat}{3}
\begin{aligned}
    &R_{2i}:\quad &&S_{2i} &&\rightarrow \quad S_{2i+1}\\
    &R_{2i+1}:\quad&&S_{2i+1} &&\rightarrow \quad S_{2i}
\end{aligned}
\end{alignat}
for $i\in\{0,\dots,N-1\}$. 
They take place at the rates explained in sec.~\ref{sec:model}. In particular, the off-rates, $k_{\text{off}}$ have to be updated in every simulation step as they change with the configuration of open and closed bonds; the vector that contains the rates is denoted by $\mathbb{K}= (k_0,....,k_{2N-1})$. Of course, reaction $R_j$ is inhibited if $X_j=0$, that is the reagent is not present at the time. This is achieved by multiplying the vector of rates, $\mathbb{K}$, component-wise with $\mathbb{X}$, giving us the vector $\mathbb{A} = (a_0,\dots,a_{2N-1}) :=(X_0\cdot k_0,\dots,X_{2N-1}\cdot k_{2N-1})$ of rates that takes the availability of reagants into consideration.

The general idea of the Gillespie algorithm is to take discrete steps, in each of which one randomly draws the next reaction and the time it takes until this reaction occurs \cite{gillespie1977exact}. 
The next reaction is chosen based on the rates by the following principle: Draw $r$ uniformly on $[0,1]$. Let $A=\sum_{j=0}^{2N-1}a_j$. Then we choose as the next reaction $R_j$ the one with
\begin{align}\label{eq:choose reaction}
    \sum_{j'=0}^{j-1}a_{j'}< r\cdot A\leq \sum_{j'=0}^ja_{j'}.
\end{align}
Note how this favours larger rates over smaller ones. The instance of the next reaction is simulated by drawing another $r'$ from a uniform distribution on $[0,1]$ and setting
\begin{align}\label{eq:reaction time step}
    \tau = \frac{1}{A} \cdot\log(1/r').
\end{align}
Upon choosing $R_j$ and $\tau$, we update the current quantities by setting $X_j=0$ and the relevant product (either $X_{j-1}$ or $X_{j+1}$) to $1$ and the time by adding $\tau$ to it.

\section{Force integration}\label{app:forceIntegral}

For the force integration in section \ref{sec:continuum_approx}, the following integral is needed:
\begin{align}\label{eq:standard integral}
    \int\frac{e^{-L/2\ell}t+e^{L/2\ell}t^{-1}}{at^2+b} dt = \frac{1}{2}\Big(\frac{e^{-L/2\ell}}{a}-\frac{e^{L/2\ell}}{b}\Big)\log(at^2+b) + \frac{e^{L/2\ell}}{b}\log(t).
\end{align}
To show this identity, we rewrite the integrand as
\begin{align}
    \frac{e^{-L/2\ell}t+e^{L/2\ell}t^{-1}}{at^2+b} 
    &= \frac{(be^{-L/2\ell}-ae^{L/2\ell})t^2+(b+at^2)e^{L/2\ell}}{bt\cdot(at^2+b)}\nonumber \\ 
    &=\frac{(be^{-L/2\ell}-ae^{L/2\ell})t}{b\cdot(at^2+b)} +\frac{e^{L/2\ell}}{bt}.
\end{align}
In this form, the function can be integrated in a straight-forward way
\begin{align}
    \int \frac{e^{-L/2\ell}t+e^{L/2\ell}t^{-1}}{at^2+b} dt 
    =\frac{1}{2}\Big(\frac{e^{-L/2\ell}}{a}-\frac{e^{L/2\ell}}{b}\Big)\log(at^2+b) +\frac{e^{L/2\ell}}{b}\log(t).   
\end{align}

Plugging the boundary values into the first logarithmic term for one of the integrals under consideration in Eq.~\eqref{eq:continuous model variable transform}, we find
\begin{align}
    \log & (a_it^2+b_i)|_{e^{x_i/\ell}}^{e^{x_{i+1/\ell}}} = \log\Big(\frac{a_ie^{2x_{i+1}/\ell}+b_i}{a_ie^{2x_i/\ell}+b_i}\Big) 
    =  \log\Big(\frac{a_ie^{x_{i+1}/\ell}+b_ie^{-x_{i+1}/\ell}}{a_ie^{x_i/\ell}+b_ie^{-x_i/\ell}}\Big)+\log\Big(e^{(x_{i+1}-x_i)/\ell}\Big). \nonumber \\
\end{align}
Inserting the expressions for $a_i$ and $b_i$ from Eq.~\eqref{eq:definition_bi} gives
\begin{align}
    \log & (a_it^2+b_i)|_{e^{x_i/\ell}}^{e^{x_{i+1/\ell}}} =\log\Bigg(\frac{\sum_{j=0}^{n-1}\cosh\big((L/2-|x_j-x_{i+1}|)/\ell\big)}{\sum_{j=0}^{n-1}\cosh\big((L/2-|x_j-x_i|)/\ell\big)}\Bigg)+(x_{i+1}-x_i)/\ell,
\end{align}
with which we obtain the force expression 
\begin{equation}\label{eq:continuous model with correction}
\begin{aligned}
    f_0 
    =\sigma \ell\sum_{i=0}^{n-1}&\frac{1}{2} \Big(\frac{e^{-L/2\ell}}{a_i}-\frac{e^{L/2\ell}}{b_i}\Big) \log\Big(\frac{\sum_{j=0}^{n-1}\cosh\big((L/2-|x_j-x_{i+1}|)/\ell\big)}{\sum_{j=0}^{n-1}\cosh\big((L/2-|x_j-x_i|)/\ell\big)}\Big)\\
    &+\frac{1}{2} \Big(\frac{e^{-L/2\ell}}{a_i}+\frac{e^{L/2\ell}}{b_i}\Big)(x_{i+1}-x_i)/\ell.
\end{aligned}
\end{equation}

\section{Derivation of rupture conditions for $\ell \rightarrow \infty$ and $\ell \rightarrow 0$}\label{sec:details glocal model}

In the following, we sketch the derivation of the rupture conditions for the global and local limit, $\ell \rightarrow \infty$ and $\ell \rightarrow 0$, respectively. For the global force distribution, this was first studied by Bell \cite{bell1978models}. Here, the critical value of $K$ as a function of $\sigma$ is derived from condition \eqref{eq:critical bond number global} for the balance of on- and off-rates. 
We simplify this condition by using the fraction of closed bonds $\Tilde{n}= n/N$ and 
$k_{\text{off},0}=k_{\text{on}}\cdot \Big(\frac{1}{K}-1\Big)$, so that 
\begin{align}\label{eq:kappa}
    \frac{1}{\Tilde{n}}-1 = \Big(\frac{1}{K}-1\Big)\cdot \exp\big(\tilde{\sigma}/\Tilde{n}\big).
\end{align}
The right-hand side of the equation is larger than the left-hand side for $\Tilde{n}=1$ as well as in the limit $\Tilde{n}\rightarrow0$.
We thus distinguish two cases depending on whether both curves have intersection points in-between. The minimum of the difference between the two curves is located at $\Tilde{n}^\ast$ such that
\begin{equation}
\label{eq:critical n} 
    1 = \tilde{\sigma}\cdot\Big(\frac{1}{K}-1\Big)\cdot \exp\big(\tilde{\sigma}/\Tilde{n}^\ast\big).
\end{equation}
An intersection points exist if, plugging $\Tilde{n}^\ast$ back into Eq.~\eqref{eq:kappa}, 
\begin{align}
    \frac{1}{\Tilde{n}^\ast}-1 \geq \frac{1}{\tilde{\sigma}},
\end{align}
which gives the condition
\begin{align}
    \tilde{\sigma}\cdot\Big(\frac{1}{K}-1\Big)\leq \exp\big(-(1+\tilde{\sigma})\big).
\end{align}
or
\begin{align} 
    K \geq 
    \frac{\tilde{\sigma}}{\exp\big(-(1+\tilde{\sigma})\big)+\tilde{\sigma}}.
\end{align}
Note that the critical value of $K$ 
is in $[0,1]$ for all $\sigma\geq 0$. For $K$ sufficiently high (i.e. sufficiently strong binding) or for $\tilde{\sigma}$ sufficiently small (small force), 
there are two solutions of Eq.~\eqref{eq:kappa}. The larger value is stable and the smaller unstable, as the total on-rate is higher than the off-rate in the interval between the two solutions and smaller outside it. Therefore the number of closed bonds will stay near the larger solution until by fluctuations, it decreases below the smaller, from where the remaining bonds rupture rapidly. We note that in the stochastic model, $n=0$ is another stable state, which is however not captured by the deterministic limit due to the divergence of the off-rate for $n\rightarrow 0$. 
On the other hand, when $K<K_c$, $n=0$ is the only stable state so that the rupture happens very quickly. 

Concerning the local model, the computation of the dimensionless critical gap size $\tilde{d_c} = d_c/\delta L$ follows from the balance condition, Eq.~\eqref{eq:condition d critical}, which we write as 
\begin{equation}
    (\tilde{d_c}-1)\cdot \exp\big(-\frac{\tilde{\sigma} \tilde{d_c}}{2}\big) = 2\Big(\frac{K}{1-K}\Big)\exp\big(\frac{\tilde{\sigma}}{2K}\big).
\end{equation}
Multiplication with $e^{\tilde{\sigma}/2}$ brings the equation into a form that is solved by the Lambert W function
\begin{equation}
    -\frac{\tilde{\sigma}}{2}(\tilde{d_c}-1)\cdot \exp\big(-\frac{\tilde{\sigma} (\tilde{d_c}-1)}{2}\big) = -\tilde{\sigma}\Big(\frac{K}{1-K}\Big)\exp\big(\frac{\tilde{\sigma}}{2}(\frac{1}{K}+1)\big)
\end{equation}
from which we obtain
\begin{equation}
    \tilde{d_c} = 1-\frac{2}{\tilde{\sigma}}W_{0/-1}\bigg(-\frac{\tilde{\sigma} K}{1-K}\exp\big(\frac{\tilde{\sigma}}{2}(\frac{1}{K}+1)\big)\bigg).
\end{equation}
Here $W_{0/-1}$ are the two branches of the Lambert $W$ function. The relevant one for crack initiation is the branch for $-1$.

\section{Force on bonds around a gap}\label{sec:details calculations force at gap}
To obtain the force on the bonds around a gap of size $d$, we consider the continuum approximation of force on a closed bond,  \eqref{eq:force_approximation}.
When there is a gap of size $d$ around the bond at $i=0$, we assume that the other closed bonds are distributed equally at $x_q = q\cdot \frac{1}{K}$ with $(n-1)\cdot \frac{1}{K}=N-\tilde{d}$, because the ratio of closed bonds to open bonds is $n/N = K$. Under this condition, the force term is
\begin{equation}
\begin{split}
     f_0 &\approx\frac{\sigma \delta L}{2K} \sum_{q=0}^{n-2} \cdot\bigg(\Big(\sum_{r=0}^qe^{-r/K\tilde{\ell}}+\sum_{r=q+1}^{n-1}e^{(N-r/K)/\tilde{\ell}}\Big)^{-1}+ \Big(\sum_{r=0}^qe^{r/K\tilde{\ell}}+\sum_{r=q+1}^{n-1}e^{-(N-r/K)/\tilde{\ell}}\Big)^{-1}\bigg)\\
    &+\frac{\sigma \delta L \tilde{d}}{2} \bigg(\Big(\sum_{r=0}^{n-1}e^{-r/K\tilde{\ell}}\Big)^{-1}+ \Big(\sum_{r=0}^{n-1}e^{r/K\tilde{\ell}}\Big)^{-1}\bigg).
\end{split}
\end{equation}
Considering the sum of a finite geometric series, $\sum_{p=0}^{n} ar^p = a \frac{1-r^n}{1-r}$, and $n\cdot \frac{1}{K} \approx N-\tilde{d}$ we obtain
\begin{equation}\label{eq:approximation force general to specific}
    \begin{split}
     f_0 &=\frac{\sigma \delta L}{2K} \sum_{q=0}^{n-2}\bigg((1-e^{-1/K\tilde{\ell}})\Big(1-e^{-(q+1)/K\tilde{\ell}}+e^{(N-(q+1)/K)/\tilde{\ell}}(1-e^{-(n-q-1)/K\tilde{\ell}})\Big)^{-1}\\
     &\quad\quad\quad+(1-e^{1/K\tilde{\ell}})\Big(1-e^{(q+1)/K\tilde{\ell}}+e^{-(N-(q+1)/K)/\tilde{\ell}}(1-e^{(n-q-1)/K\tilde{\ell}})\Big)^{-1}\bigg)\\
    &+\frac{\sigma \delta L \tilde{d}}{2} \bigg(\frac{1-e^{-1/K\tilde{\ell}}}{1-e^{-n/K\tilde{\ell}}}+ \frac{1-e^{1/K\tilde{\ell}}}{1-e^{n/K\tilde{\ell}}}\bigg)\\
    &\approx \frac{\sigma \delta L}{2K} \sum_{q=0}^{n-2} \bigg((1-e^{-1/K\tilde{\ell}})\Big(1-e^{-(q+1)/K\tilde{\ell}}+e^{(N-(q+1)/K)/\tilde{\ell}}-e^{\tilde{d}/\tilde{\ell}}\Big)^{-1}\\
    &\quad\quad\quad+(1-e^{1/K\tilde{\ell}})\Big(1-e^{(q+1)/K\tilde{\ell}}+e^{-(N-(q+1)/K)/\tilde{\ell}}-e^{-\tilde{d}/\tilde{\ell}}\Big)^{-1}\bigg)\\
    &+\frac{\sigma \delta L \tilde{d}}{2} \bigg(\frac{1-e^{-1/K\tilde{\ell}}}{1-e^{-(N-\tilde{d})/\tilde{\ell}}}+ \frac{1-e^{1/K\tilde{\ell}}}{1-e^{(N-\tilde{d})/\tilde{\ell}}}\bigg).
\end{split}
\end{equation}

\section{Values of the critical gap size}\label{sec:relative error critical gap size}
An overview of the different estimates of the critical gap size as obtained by the different methods is given in Table~\ref{tab:comparison_good_approximation} for two sets of parameters. We see that the results of the three estimation methods, using the local model, Eq.~\eqref{eq:critical d local}, using the small $\tilde{\ell}$ approximation, Eq.~\eqref{eq:critical_gap_size_small_l}, and using the continuum approximation, Eq.~\eqref{eq:approximation force specific}, are basically the same for $\tilde{\ell}=0.2$, for which the local model provides a very good approximation. 
For $\tilde{\ell}=0.5$, however, we see that the estimate using the continuum approximation is much closer to the simulation result than those based on the local model and the small-$\tilde\ell$ approximation. We consider this to be a significant improvement compared to Eq.~\eqref{eq:critical_gap_size_small_l}; however at the price of a more complicated and computationally demanding expression. The main advantage of Eq.~\eqref{eq:approximation force specific} is that it allows us to compute the force at the largest gap with good accuracy and thereby approximate the critical gap size numerically, based solely on the parameters $K,N,\sigma$ and $\tilde{\ell}$ and not the entire bond configuration. 

\begin{table}[tb]
\centering
\caption[Different estimates of critical gap size]{Estimates of the critical gap size $\tilde{d_c}$ for $N=200,K=0.9,\tilde{\sigma}=0.7$ and $\tilde{\ell}=0.2$, resp. $0.5$. The estimate using the continuum approximation Eq.~\eqref{eq:approximation force general} is clearly closer to the value obtained in simulations than the one from the local model. Figure~\ref{fig:critical gap size small l} contains a visualisation of the situation.}
\label{tab:comparison_good_approximation}
\begin{tabular}{|c||c|c|c|c|}
\hline Estimation method & local model & small-$\tilde{\ell}$ appr. & continuun appr. & simulations\\ \hline\hline
 $\tilde{\ell}=0.2$    & $9.66$ & $9.663$ &  9.71 & $9.6\pm2.5$ \\
 $\tilde{\ell}=0.5$   & $9.66$ &$9.5$ & 11.19 & $10.8\pm2.6$ \\\hline
\end{tabular}
\end{table}


%

\end{document}